\documentclass[onecolumn,a4paper]{IEEEtran}
\usepackage{geometry}
\ifCLASSINFOpdf
\else
   \usepackage[dvips]{graphicx}
\fi
\usepackage{url}

\usepackage{amsmath}
\usepackage{amsfonts}
\usepackage{epsfig}
\usepackage[colorinlistoftodos]{todonotes}
\usepackage[colorlinks, linkcolor=red, citecolor=blue]{hyperref}
\usepackage{amssymb}
\usepackage[noend]{algpseudocode}
\usepackage{algorithmicx,algorithm}
\usepackage{subfig}

\begin{document}

\title{Sliding Differential Evolution Scheduling for Federated Learning in Bandwidth-Limited Networks}

\author{Yifan Luo, Jindan Xu, \IEEEmembership{Student Member, IEEE}, Wei Xu, \IEEEmembership{Senior Member, IEEE},\\ Kezhi Wang, \IEEEmembership{Senior Member, IEEE}

\thanks{Y. Luo, J. Xu are with the National Mobile Communications Research Laboratory (NCRL), Southeast University, Nanjing 210096, China (email: \{213161316, jdxu\}@seu.edu.cn).}
\thanks{W. Xu is with the National Mobile Communications Research Lab, Southeast University, Nanjing 210096, China, and also with Purple Mountain Laboratories, Nanjing 211111, China (wxu@seu.edu.cn).}
\thanks{K. Wang is with Department of Computer and Information Sciences, Northumbria University, Newcastle upon Tyne NE1 8ST, U.K. (e-mail: kezhi.wang@northumbria.ac.uk).}}

\maketitle

\begin{abstract}
Federated learning (FL) in a bandwidth-limited network with energy-limited user equipments (UEs) is under-explored.
In this paper, to jointly save energy consumed by the battery-limited UEs and accelerate the convergence of the global model in FL for the bandwidth-limited network, we propose the sliding differential evolution-based scheduling (SDES) policy.
To this end, we first formulate an optimization that aims to minimize a weighted sum of energy consumption and model training convergence.
Then, we apply the SDES with parallel differential evolution (DE) operations in several small-scale windows, to address the above proposed problem effectively.
Compared with existing scheduling policies, the proposed SDES performs well in reducing energy consumption and the model convergence with lower computational complexity.
\end{abstract}


\IEEEpeerreviewmaketitle

\section{Introduction}
\IEEEPARstart{I}{n} future wireless networks, by building and utilizing the  computation capability in edge nodes, e.g., access points (APs), edge networks can be established and are able to conduct complex task via
intelligent scheduling and processing \cite{park2019wireless}. Several works utilizing machine learning have been proposed for future communications, e.g.,
C-RAN \cite{luo2020power}, MIMO channel information feedback system \cite{lu2018mimo} and multi-antenna quantization \cite{8845636}.
However, massive raw data generated by user devices triggers two key problems, i.e., privacy disclosure and high cost from data transmission, making the above-mentioned intelligent applications, difficult to process in wireless networks. 
Federated learning (FL) has been proposed by Google, as a promising machine learning (ML) technology to solve the above problems   \cite{mcmahan2016communication}. 

Specifically, there are two key challenges for the deployment of FL in wireless networks. On one hand, local data samples in UEs are diversely distributed, i.e., non-independent and identically distributed (non-IID) and unbalanced \cite{mcmahan2016communication,zhao2018federated}. To strike a balance between computational efficiency and convergence, Google proposed a novel FL architecture, referred to as FedAvg \cite{mcmahan2016communication}. Additionally, other researchers, e.g. \cite{FEDL}, tried to accelerate the convergence by converting the optimization problem into sub-problems.
Another challenge is that the limited capacity of communications, e.g., limited bandwidth.
To address it, scheduling policy-aided FL architectures were utilized in \cite{yang2019scheduling, zeng2019energy, AoU}. The authors in \cite{AoU} adopted the age of update (AoU) as the scheduling policy to accelerate model convergence in mobile edge networks. Also, the authors in \cite{zeng2019energy} proposed a bandwidth resource scheduling policy for FL in wireless networks. 
Their scheduling policies only take FL model convergence into consideration \cite{AoU} or just adopt randomly selection \cite{yang2019scheduling}.

However, there is little literature related to joint energy efficiency and model convergence for federated learning in wireless communication. The authors in \cite{FEDL} carefully analysed the trade-off between energy consumed by UEs and FL convergence with no bandwidth-limited constrains. The authors in \cite{zeng2019energy} proposed an energy-efficient bandwidth resource scheduling policy for FL in wireless networks, and it only considered the communication energy cost.

Against the above background, in this paper, we aim to save the energy consumption of the UEs and improve the convergence performance of FL in a bandwidth-limited network.
We propose an efficient sliding differential evolution-based scheduling (SDES) with lower computational complexity compared with the existing methods.
To the best of our knowledge, it is the first time to solve the above problem well. 
In detail, we introduce a convergence reference (CR) of the overall training model and propose the SDES policy to reduce the energy consumption and accelerating model convergence, by choosing the optimal subgroup of the UEs.
Compared to conventional mathematical iterative tools, the proposed SDES can process the computational tasks in a parallel model.
Experiment verifies the effectiveness of the proposed solution, in terms of both energy saving and convergence acceleration in the bandwidth-limited network.

\section{System Model}

\begin{figure}[t]
   \centering
   \includegraphics[height=5cm]{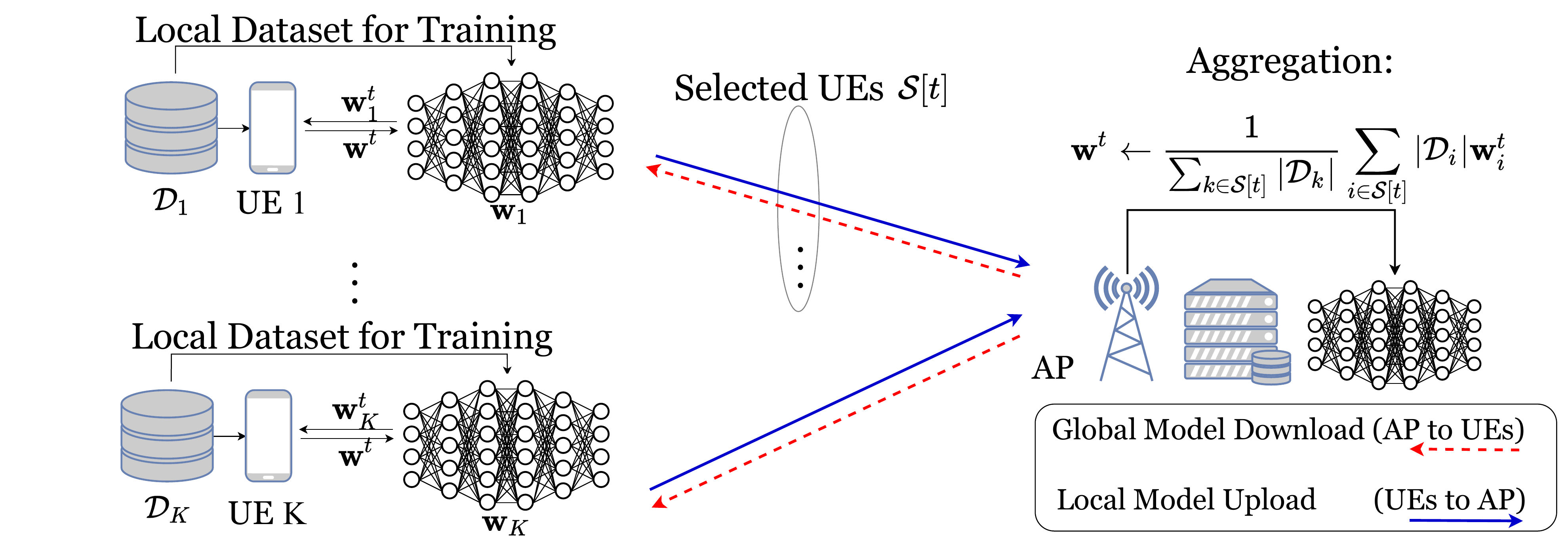} \setlength\belowcaptionskip{0cm}
   \caption{Federatd learning in wireless networks}
   \label{WFL}
\end{figure}

We consider a FL system as shown in Fig.~\ref{WFL}, where a set $\mathcal{K}$ of $K$ UEs are connected to one AP. Each UE $k$ stores a local dataset $\mathcal{D}_k$, with its size denoted by $D_k = |\mathcal{D}_k|$. Thus, the whole data size equals to $D=\sum_{k=1}^{K}{D_k}$. Dataset $\mathcal{D}_k$ denotes the collection of data samples in the form of input-output pairs as $\{\mathbf{x}_i^{(k)},y_i^{(k)}\}^{D_k}_{i=1}$, where $\mathbf{x}_i^{(k)} \in \mathbb{R}^d$ is an input sample vector with $d$ features, and $y_i^{(k)} \in \mathbb{R}$ is the labeled output value for sample $\mathbf{x}_i^{(k)}$. Considering the user preference, different $\mathcal{D}_k$'s are non-IID and their corresponding data size $D_k$ varies.

\subsection{Model Convergence}

The goal of AP is to learn a statistical model over the data that resides on the $K$ associated UEs. Mathematically, AP needs to fit the model parameter $\mathbf{w}^t \in \mathbb{R}^d$ which characterizes the output $y_i$, by minimizing a particular loss function $f_i(\mathbf{w}^t)=\ell(\mathbf{x}_i^{(k)},y_i^{(k)}; \mathbf{w}^t)$ in the $t$-th communication round. Formally, the loss function on the dataset of UE $k$ is
\begin{equation}
    F_k(\mathbf{w}^t):= \frac{1}{D_k}\sum_{i \in \mathcal{D}_k} 
    f_i(\mathbf{w}^t).
\label{perloss}
\end{equation}

Then, the global loss function minimization problem in AP can be expressed as
\begin{equation}
   \label{global_loss_function} \min \limits_{\mathbf{w}^t} \quad F(\mathbf{w}^t):=\sum_{k=1}^{K}\frac{D_k}{D}{ F_k(\mathbf{w}^t)}.
\end{equation} 

To protect the user privacy, UE $k$ only exchanges its model parameters $\mathbf{w}^t_k$ with AP. 

\subsection{Energy Consumption}
In bandwidth-limited systems, the number of UEs, $K$, far exceeds the number of subchannels, $N$. Only a small portion of UEs, referred to as the updating set $\mathcal{S}[t]$, are selected for participating in the $t$-th communication round. $\mathcal{S}[t]=\{k \mid S_k[t]=1, k=1,2,...,K\}$, where $S_k[t]=1$ implies that UE $k$ is in the updating set $\mathcal{S}[t]$, otherwise $S_k[t]=0$.
The energy consumed by UEs in the updating set $\mathcal{S}[t]$ consists of two componemts, i.e., transmitting energy consumption and computing energy consumption. 

In fact, all UEs share the same model with their local parameters, and we 
use constant $J$ to denote its size of $\mathbf{w}^t$. We assume in the $t$-th communication round, UE $k$ is assigned with the $n$-th subchannel with channel gain $h_{k,n}$ and bandwidth $B_n$. Then, the achievable data rate of UE $k$ can be
\begin{equation}
    r_k=B_n\ln{\left( 1+\frac{h_{k,n}^2 P_{k,n}}{N_0} \right)},
\label{trate}
\end{equation}
where $P_{k,n}$ denotes the corresponding power allocation, and $N_0$ denotes the variance of the white Gaussian noise.

To obtain minimal transimit power, we assume the achievable rate $r_k$ equals to the threshold transmission rate $R_k$, and the energy for signal transmission for UE $k$ is formulated as
\begin{equation}
    E_{k,\text{TP}}= \tau_k \cdot P_{k,n} = \frac{J}{R_k} \cdot \frac{N_0}{h_{k,n}^2}\left(e^{\frac{R_k}{B_n}}-1\right), \label{tp} 
\end{equation}
where $\tau_k$ is time duration of the signal transmission process.

On the other hand, the computing energy consumed by UE $k$ to train its local model can be written as
\begin{equation}
   E_{k,\text{CP}}= \sum_{i=1}^{c_k D_k}\frac{\alpha_k}{2}{f_k}^2=\frac{\alpha_k}{2}{c_k}{D_k}{f_k}^2,
\label{cp} 
\end{equation}
where $c_k$ denotes the number of CPU cycles for executing one sample of data;  $c_k D_k$ denotes the number of CPU cycles in one local round; and $f_k$ is the CPU-cycle frequency. Then, the total energy consumption of UEs in the $t$-th communication round is
\begin{equation}
    E_{\text{P}}[t]= \sum_{k=1}^{K} {S_k[t] (E_{k,\text{TP}}+\kappa E_{k,\text{CP}})}, \quad S_k[t] \in \{0,1\}
\label{EP} 
\end{equation}
where $\kappa$ is the number of local rounds for local model training.

\subsection{Problem Formulation}
We aim to minimize the weighted sum of global loss function in \eqref{global_loss_function} and the total energy consumption in \eqref{EP} in the $t$-th communication round as:
\begin{subequations}\label{ooptimfun}
   \begin{align}
   \label{ooptim1}\min \limits_{\mathcal{S}[t]} \quad & F(\mathbf{w}^t) + \zeta E_{\text{P}}[t], \ t \in \{0,1,..,T\} & \\
   \label{ooptim2}\mathrm{s.t.} \quad & S_k[t] \in \{0,1\}, \ \forall k \in \{1,2,\cdots,K\} & \\
   \label{ooptim3}& \sum_{k=1}^{K} S_k[t] = N, &
   \end{align}
\end{subequations}
where $\zeta$ is the factor to balance the loss function and the energy consumption; $T$ denotes the number of communication rounds between AP and UEs;
constraint \eqref{ooptim3} shows that the number of the available sub-channels is $N$.



\section{Sliding Differential Evolution Based Scheduling}
There are two challenges for solving the optimization problem in \eqref{ooptimfun}.
On one hand, due to the limited bandwidth, only a small portion of UEs' training loss and model parameters can be updated to AP. This makes it impossible to calculate the global loss $F(\mathbf{w}^t)$ in \eqref{ooptim1} accurately.
Also, it is difficult to get the relationship between $\mathcal{S}[t]$ and $\mathbf{w}^t$ in \eqref{ooptimfun}, where $\mathbf{w}^t$ relies on model training. 
On the other hand, this optimization problem is a combinatorial problem which does not normally have low complexity solutions. 
For instance, in the case of 100 UEs and 25 available sub-channels, $C(100,25) \approx 2.4\times 10^{23}$ searches are needed in the exhaustive searching, where $C(n,k) = \frac{n!}{k!(n-k)!}$ is the searching number of $k$-combinations from a given set of $n$ elements. 

In this section, we first introduce the convergence reference (CR) function to replace $F(\mathbf{w}^t)$ for model convergence above, where two parameters in \eqref{ooptim1} are unified into one set, $\mathcal{S}[t]$. 
Based on CR value and energy consumption expression, we propose the SDES policy, for solving \eqref{ooptimfun} efficiently.


\subsection{Convergence Reference (CR) Function}
To solve the aforementioned two challenges, we propose the concept of convergence reference (CR). In CR function, we collect the useful data from the updated UEs for model convergence, and utilize CR function to improve the model convergence based on these data. The CR function transforms the convergence problem into finding the optimal updated set $\mathcal{S}[t]$, and it is consistent with the energy problem in \eqref{ooptim1}.

We introduce CR function based on staleness-loss (SL) measure for convergence, where CR value is utilized to select the optimal subgroup of users. We re-formulate the convergence performance of local models into SL measure based on two existing methods, i.e., staleness and training loss method.

The staleness method, also referred to as AoU \cite{AoU} can transform convergence value into model training times. It records the duration ${T}_k[t]$ 
which the UE $k$ uploading its model in the $t$-th round,
written as ${T}_k[t]=({T}_k[t-1]+1)(1-S_k[t-1])$. Staleness method leverages ${T}_k[t]$ to avoid over-training and under-training \cite{chen2019deep}.

Moreover, the training loss method records the training loss for all the UEs, where the training loss for UE $k$ is $\mathcal{L}_k[t] = F_k(\mathbf{w}^t)$ in the $t$-th communication round in \eqref{perloss}.

Based the above two methods, we introduce staleness-loss (SL) measures, by combining the staleness and training loss. The set of SL value for all the UEs in the $t$-th communication round can be written as:
\begin{equation*}
   \mathcal{C}[t]=\{C_1[t],C_2[t],\cdots,C_k[t],\cdots, C_K[t]\}
\end{equation*}
where $C_k[t] = T_k[t] \mathcal{L}_k[t]$.

Then, we introduce the convergence reference (CR) function based on SL value as
\begin{equation}
    T_\text{L}[t]=
    \frac{(\sum^{K}_{k=1}{D_k V_k[t]  S_k[t]})^{1-\beta}}{1-\beta},
    \label{Combined_Cov}
\end{equation}
where $\beta \in (0,1)$ is a constant to adjust the sensitivity to the value change of $\sum^{K}_{k=1}{D_k  V_k[t]  S_k[t]}$, and $D_k$ is the size of user data. $V_k[t] \in \{T_k[t], \mathcal{L}_k[t], C_k[t]\}$ denotes the method used in CR function. Then, in the $t$-th communication round, the objective in \eqref{ooptim1} can be re-written as:

\begin{subequations}\label{optimfun}
   \begin{align}
   \label{optim1} \min \limits_{\mathcal{S}[t]} \quad & - T_{\text{L}}[t] + \zeta E_{\text{P}}[t], \ t \in \{0,1,..,T\} & \\
   & (7b), \ (7c) \nonumber  &
   \end{align}
\end{subequations}

\subsection{The Sliding Differential Evolution (SDE) Concept}
The optimization problem in \eqref{optimfun} is NP-hard. Differential evolution (DE) \cite{storn1996usage} is a common method to solve the above kind of problem, where DE generates $M$ individuals with the chromosome scale, $K$, i.e., the dimensionality of \eqref{optim1}, for each individual. We assume there are $G_{\text{DE}}$ generations in DE, and DE algorithm terminates after exceeding $G_{\text{DE}}$ iterations. The execution time is proportional to the objective function evaluation $c(K)$ of \eqref{optimfun} \cite{opara2019differential} with $K$ dimensionality, and the number of elementary operations is proportional to the maximal iteration number $G_{\text{DE}}$ and the population size, i.e., $\mathcal{O}(c(K) \cdot M \cdot G_{\text{DE}})$. However, traditional DE methods suffer from heavily computational complexity when the scale of \eqref{optimfun} increases. Therefore, we propose the concept of sliding differential evolution (SDE) to decrease the computational complexity by reducing the scale of chromosomes from $K$ to $W$, and the number of generations from $G_{\text{DE}}$ to $G_{\text{SDES}}$ with parallel computation, where $W$ is the length of energy windows and $G_{\text{SDES}}$ is the number of generations in SDE. Consequently, its complexity can be given by $\mathcal{O}(c(W) \cdot M \cdot G_{\text{SDES}}).$

\subsection{Sliding Differential Evolution-based Scheduling (SDES)}
\begin{figure}[t]
	\centering
	\includegraphics[height=8cm]{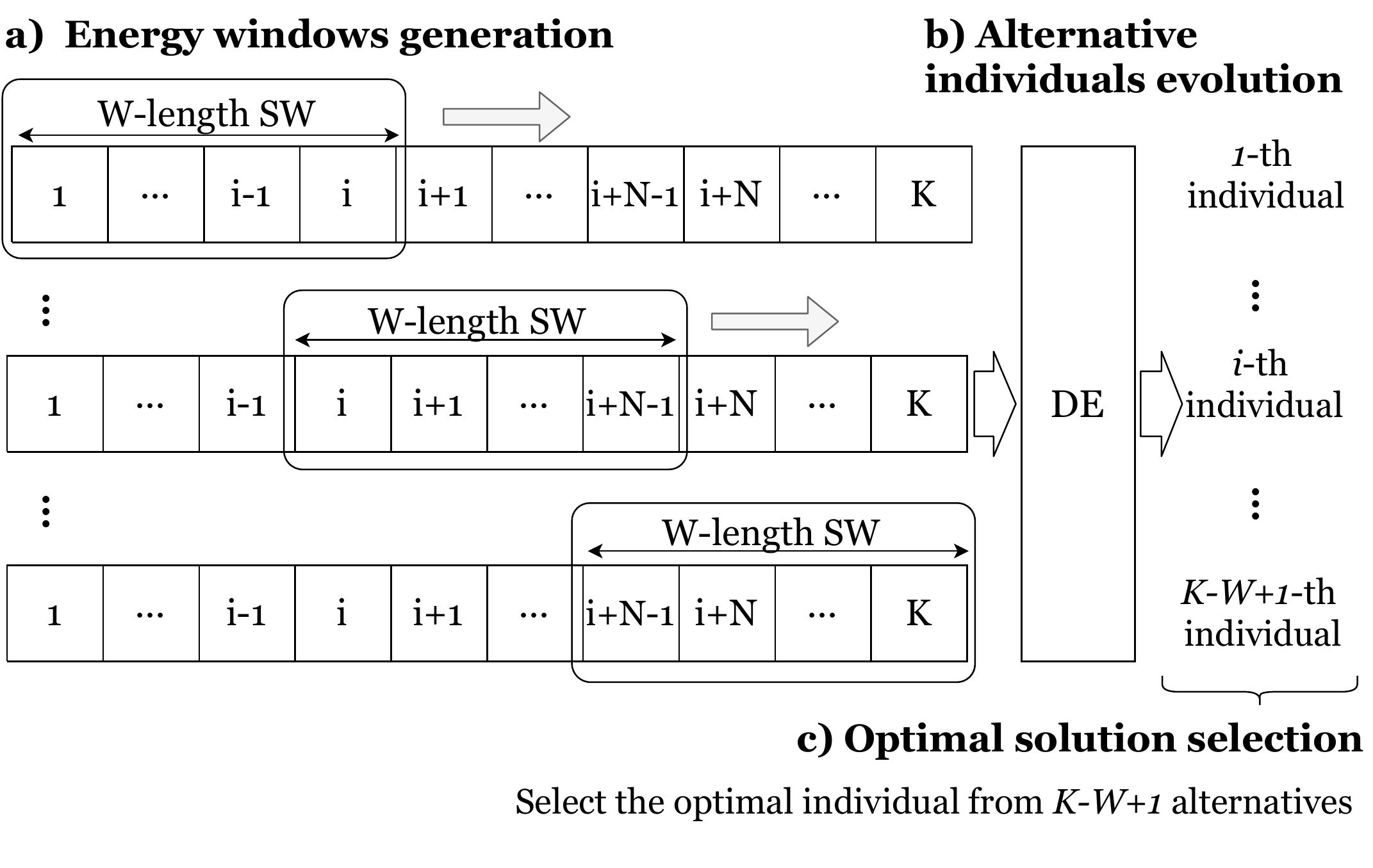}
	\caption{Sliding differential evolution algorithm}
	\label{SWpic}
\end{figure}

The sliding differential evolution-based scheduling (SDES) algorithm is shown in Algorithm~\ref{alg:SDES} and its process is summarized in Fig.~\ref{SWpic}. SDES takes the steps as follows:
\begin{itemize}
   \item {\bf a)} Energy windows generation: 
   We first leverage $W$-length SW to generate $K-W+1$ energy windows, where $N \le W \le K$. Specifically, we first sort all UEs according to their energy consumption from small to large, and then align the head of the SW with the first UE to select the first $W$ UEs in one window. 
   Similarly, we slide the window to the end of the queue for another $K-W$ energy windows. 
   \item {\bf b)} Alternative individuals evolution:
   In each window, we utilize DE to evolve one alternative individual, i.e., one scheduling scheme with minimal value of \eqref{optim1} from the $W$ UEs, and the scale of chromosome scale is $W$. We conduct DE operations in $K-W+1$ energy windows parallelly, and generate $K-W+1$ alternative individuals.
   \item {\bf c)} Optimal solution selection:
   We select the optimal individual, i.e., the best solution of \eqref{optim1} from the $K-W+1$ alternative individuals. 
\end{itemize}

\begin{algorithm}[t]
\caption{ SDES }
\label{alg:SDES}
{\bf Parameters:}
$W$: the length of SW of \eqref{optim1}; $M$: the size of populations; $f_{CR}$: the crossover rate; $F$: the selection weighting factor; $G_{\text{SDES}}$: the number of generations\\
{\bf Input:}
Optimization problem \eqref{optim1}; $K-W+1$ energy windows
\begin{algorithmic}[1]
   \For{$w=1,2,\cdots,K-W+1$} \qquad \%\% {\textbf{Step a}}
   \newline
\qquad \textit{Initialization:}
      \State Generate Initial population $P^{0,w}$ with $M$ individuals
      \For{$g=0,1,\cdots,G-1$}  \qquad \quad \%\% {\textbf{Step b}}
      \State Calculate the fitness value $Q(\cdot)$ of Generation $p^{g,w}$
         \For{each individual $\mathbf{x}^{(g,w)}_i$ in Generation $p^{g,w}$}
         \newline
\qquad \qquad \textit{(a) Mutation:}
            \State Select three individuals $r1\neq r2\neq r3$ via RWS
            \State $\mathbf{v}^{(g,w)}_i = \mathbf{x}^{(g,w)}_{r1}+F \cdot (\mathbf{x}^{(g,w)}_{r2} - \mathbf{x}^{(g,w)}_{r3}) $
            \newline
\qquad \qquad \textit{(b) Crossover:}
            \State $\mathbf{u}^{(g,w)}_i=\mathbf{x}^{(g,w)}_i$
            \State $j$ randomly selected from $\{1,2,...,D\}$, $L=1$
            \Repeat
               \State $u^{(g,w)}_{i,j} =v^{(g,w)}_{i,j}$
               \State $j = (j+1)$ modulo $K$
               \State $L = L + 1$
            \Until{$rand(0,1)<f_{CR}$ and $L<D$}
            \newline
\qquad \qquad \textit{(c) Selection:}
            \If{$Q(\mathbf{u}^{(g,w)}_i) \ge Q(\mathbf{x}^{(g,w)}_i)$}
               \State add $\mathbf{u}^{(g,w)}_i$ in the next generation $P^{g+1,w}$
            \Else{ add $\mathbf{x}^{(g,w)}_i$ in the next generation $P^{g+1,w}$} 
            \EndIf
         \EndFor
      \EndFor
   \EndFor
   \State Add the best individual in the population $P^{G,w}$ into the alternative individual list
\end{algorithmic}
{\bf Output:}
The optimal individual from $K-W+1$ alternative individuals 
\qquad \qquad \qquad \qquad \qquad \qquad \quad \%\% \textit{\textbf{Step c}}
\end{algorithm} 

For DE operations in each energy window in the above {\bf Step b)}, we define the number of generations as $G_{\text{SDES}}=\min\{\lceil \frac{C(K,W)}{M} \rceil, G_{\text{DE}} \}$, where $M$ is the number of individuals and $G_{\text{DE}}$ is the number of evolution generations in the traditional DE algorithm. Each individual $\mathbf{x}^{(g,w)}_{i}$ represents one solution of \eqref{optim1}. For instance, in the $w$-th energy window, the agent first generates the initial population $P^{0,w}=\{\mathbf{x}^{(0,w)}_1, \mathbf{x}^{(0,w)}_2,...,\mathbf{x}^{(0,w)}_M\}$. $\mathbf{x}^{(0,w)}_i=\{x^{(0,w)}_{i,1},x^{(0,w)}_{i,2},...,x^{(0,w)}_{i,W}\}$ meets the constrains of $S_k[t]$ in \eqref{ooptimfun}, where \eqref{ooptim3} is rewritten as $\sum_{j=1}^{W} x^{(0,w)}_{i,j} = N$. Any individual violating the constraint of \eqref{ooptimfun} is abandoned. Then each individual $\mathbf{x}^{(g,w)}_i$ from the $g$-th generation in the the $w$-th energy window generates the offspring with three process, given as
\begin{itemize}
   \item {\bf Mutation:} We choose three individuals from $P^{g,w}$ via roulette wheel selection (RWS) to generate $\mathbf{v}^{(g,w)}_i$.
   \item {\bf Crossover:} We cross the current individual $\mathbf{x}^{(g,w)}_i$ with $\mathbf{v}^{(g,w)}_i$ and then generate $\mathbf{u}^{(g,w)}_i$.
   \item {\bf Selection:} We choose the appropriate offspring between $\mathbf{u}^{(g,w)}_i$ and $\mathbf{v}^{(g,w)}_i$ by comparing their  fitness value.
\end{itemize}
The RWS in the process of mutation associates the probability of selecting individual $x$ with the fitness function, as $p(x)={\frac {Q(x)}{\Sigma _{j=1}^{M}Q(j)}}$.

The fitness function $Q(\cdot)$ is transformed from the optimization objective in \eqref{optim1} via linear scaling as follows
\begin{subequations}\label{DEfunc}
   \begin{align}
   \label{DEfunc1} & Q(\mathbf{x}^{(g,w)}_{i})=\alpha_1 \cdot O(\mathbf{x}^{(g,w)}_{i})+ \beta_1 &  \\
   \label{DEfunc2} \text{where} \ \ & O(\mathbf{x}^{(g,w)}_{i}) = -(-T_{\text{L}}[t] +\zeta E_{\text{P}}[t])\bigg|_{S_k[t] \in \mathbf{x}^{(g,w)}_{i}} & \\
   \label{DEfunc3} &\alpha_1 = \frac{O_{\text{avg}}}{O_{\text{avg}}-O_{\text{min}}}, \ O_{\text{avg}} = \frac{1}{M} \sum_{j=1}^{M} {\mathbf{x}^{(g,w)}_{j}} & \\
   \label{DEfunc4} & \beta_1 = \frac{-O_{\text{min}}O_{\text{avg}}}{O_{\text{avg}}-O_{\text{min}}}, \ O_{\text{min}} = \min \limits_{j=1,...,M}  {\mathbf{x}^{(g,w)}_{j}}&
   \end{align}
\end{subequations}
and \eqref{DEfunc2} takes the reverse direction of \eqref{optim1} for minimization.

\subsection{Two Cases of SDES: $W$=$K$ and $W$=$N$}
The computational complexity of DE algorithm is $\mathcal{O}(c(K) \cdot M \cdot G_{\text{DE}})$, while the one for SDES is $\mathcal{O}(c(W) \cdot M \cdot G_{\text{SDES}}).$ where $G_{\text{SDES}}=\min\{\lceil \frac{C(K,W)}{M} \rceil, G_{\text{DE}} \}$. Considering $W \le K$ and $G_{\text{SDES}} \le G_{\text{DE}}$ where two equations all reach only if $W=K$, SDES can decrease the computational complexity compared with DE.
However, the performance of scheduling policy generated by SDES is decreased when $W$ reduces.

To investigate the stability of SDES, we analyse two cases of SDES, i.e., $W$=$K$ and $W$=$N$. More specifically, when $W=K$, there is only one energy windows in SDES, and the SDES algorithm can generate the best solution of \eqref{optim1} at the highest computational cost. When $W$=$N$, all the UEs in one energy window are selected as the scheduling policy, and there is no need to generate policies by DE, where SDES generates the worst solution with the lowest computational cost.

\section{Simulation Results}


\begin{table}[t]
\begin{center}  
\caption{Parameter Settings}
\label{tal1}
\begin{tabular}{|p{50pt}|p{200pt}|p{150pt}|}  
\hline  
Symbol & Parameters & Value \\ \hline
$N$, $K$ & Number of subchannels, UEs & 25, 100 \\ \hline 
$N_0$, $B$, $R$ & Noise, bandwidth, threshold transmission rate& $10^{-8}$ W, 10Mbps, 500Kbps \\ \hline 
$\eta$, $\text{J}$, $\text{D}$ & Learning rate, model and data size & 0.1, 86.6 KB, 47.04 MB \\ \hline
$\text{f}$, $\alpha$, $\text{C}$ & CPU frequency, capacitance coefficient, cycles to execute & 2GHz, $2*10^{-28}$, 20 cycle/bit\\ \hline
$L(d_{k,n})$ & Path loss of Rayleigh fading & $99.3+20\log{d_{k,n}}$ \\ \hline  
$d_{k,n}$ & Distribution of UE $k$ & Uniform in [5,50] m \\ \hline
\end{tabular}  
\end{center}  
\end{table}

In this simulation, we adopt the orthogonal frequency division multiple access (OFDMA) system, 
and the details of the system are summarized in Table \ref{tal1}.
We assume all $N$ sub-channels share the same bandwidth of $\frac{B}{N}$. 

In FL training, the task is to classify handwritten digits using the MNIST dataset. In detail, the dataset distribution over UEs are unbalanced and non-i.i.d, where the unbalanced feature means that the dataset size varies greatly between different UEs.
The training model is a 6-layer convolutional neural network (CNN), consisting of two 5$\times$5 convolution layers with rectified linear unit (ReLU) activation. 
The two convolution layers have 10 and 20 channels respectively, and each layer has 2$\times$2 max pooling, a fully-connected layer with 50 units and ReLU activation, and a log-softmax output layer. 

Next, we validate the overall performance of SDES with the respect of energy saving and model convergence, through CR function in \eqref{Combined_Cov}, where SDES ($W$=$K$) and SDES ($W$=$N$) are examined. The measure $V_k[t]$ in CR function can be selected from $\{T_k[t], \mathcal{L}_k[t],C_k[t]\}$.
We adopt the FedAvg from Google \cite{mcmahan2016communication} as the benchmark and set $\zeta=5$. 
In detail, when the weight factor $\zeta > 5$, SDES will focus more on energy saving, and consequently improve energy efficiency, however, at the expense of worse convergence performance. When $\zeta < 5$, the model convergence gets improved, and the performance of energy efficiency will deteriorate. 

\begin{figure}[t]
	\centering
	\subfloat[Training loss ($W$=$N$)]{
		\begin{minipage}[t]{0.5\linewidth}
			\centering
         \includegraphics[width=3 in]{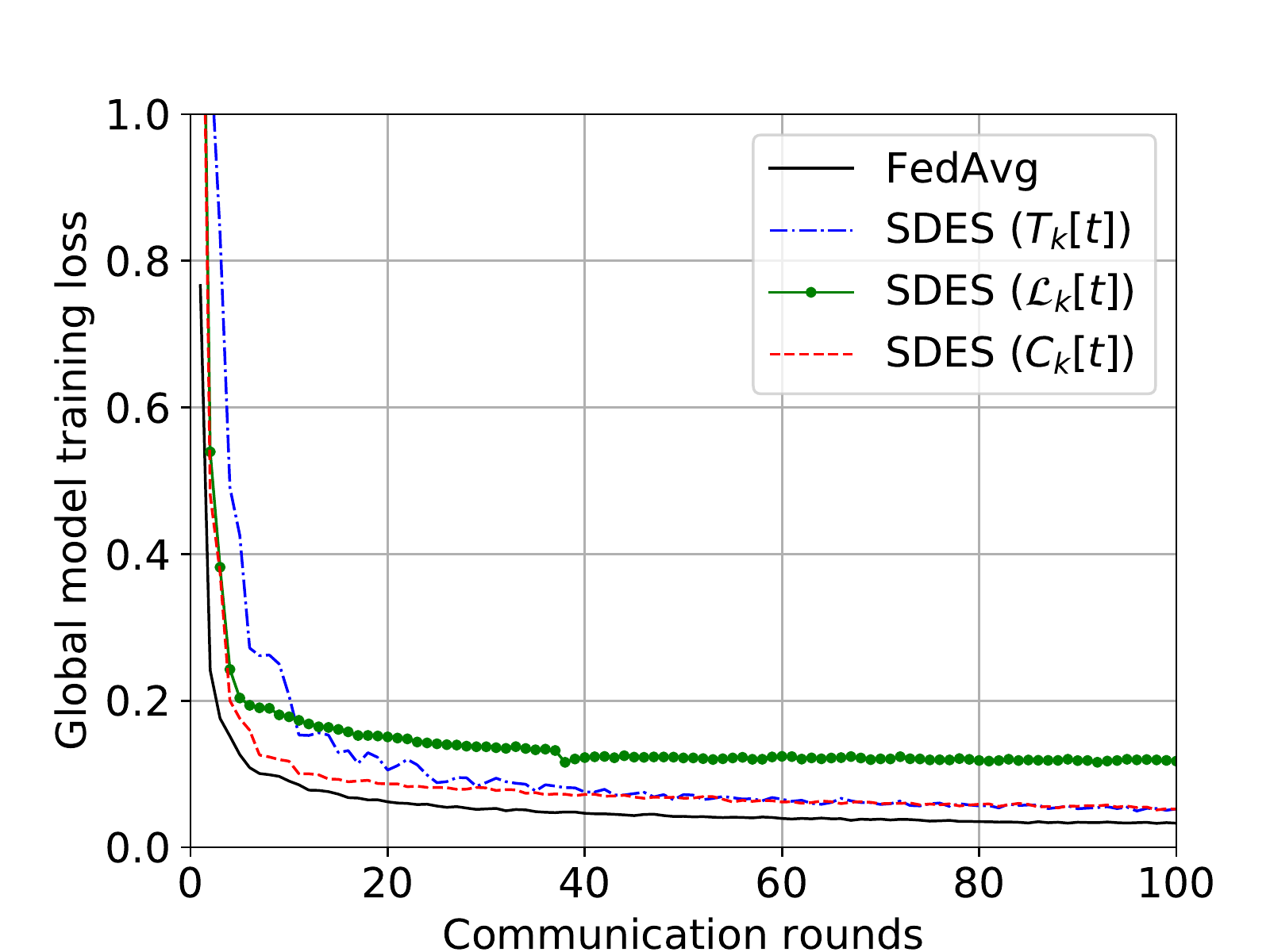}
   \end{minipage}} %
   \subfloat[Training loss ($W$=$K$)]{
		\begin{minipage}[t]{0.5\linewidth}
			\centering
			\includegraphics[width=3 in]{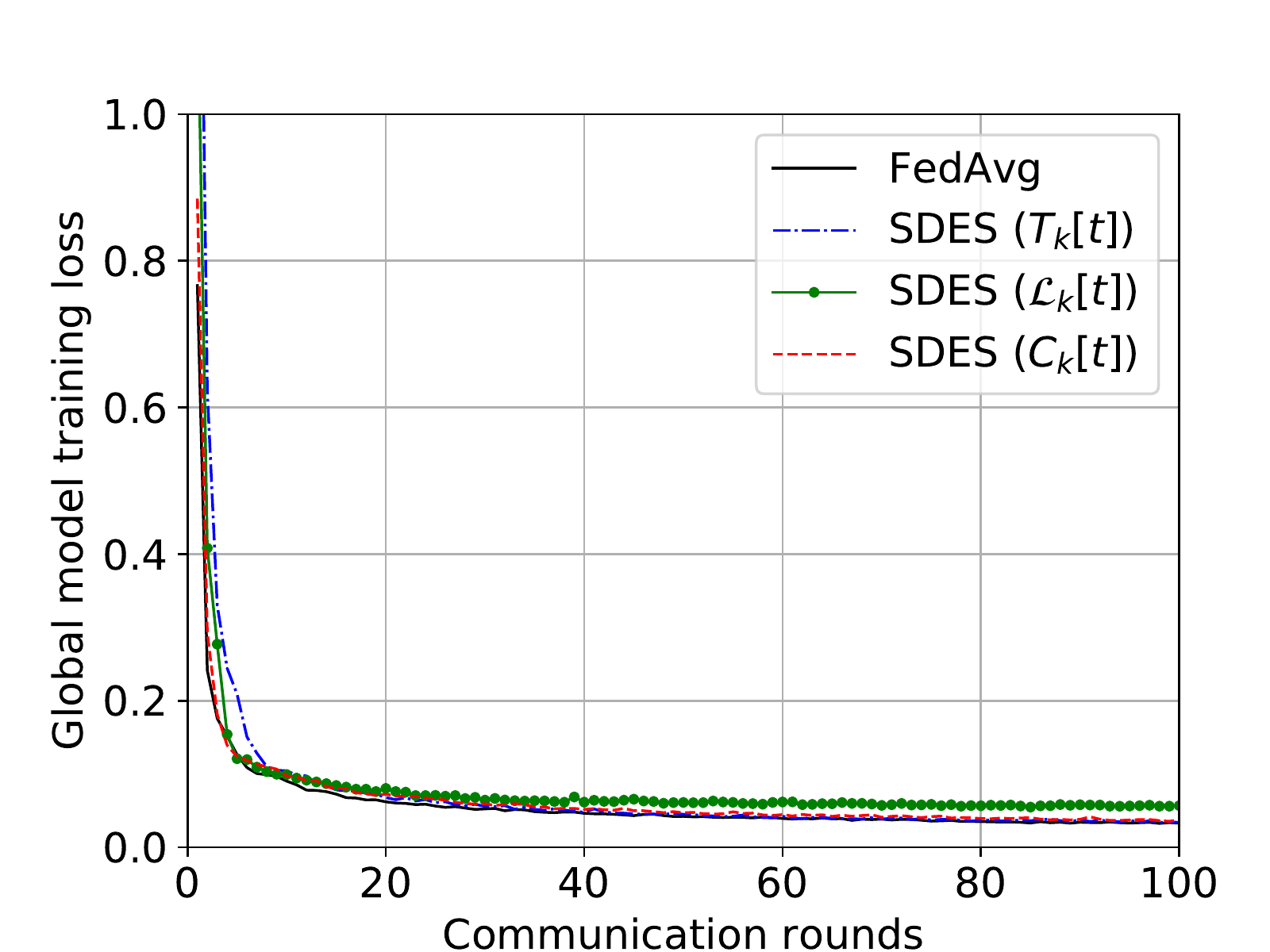}
   \end{minipage}}
	\\
	\subfloat[Energy consumption ($W$=$N$)]{
		\begin{minipage}[t]{0.5\linewidth}
			\centering
			\includegraphics[width=3 in]{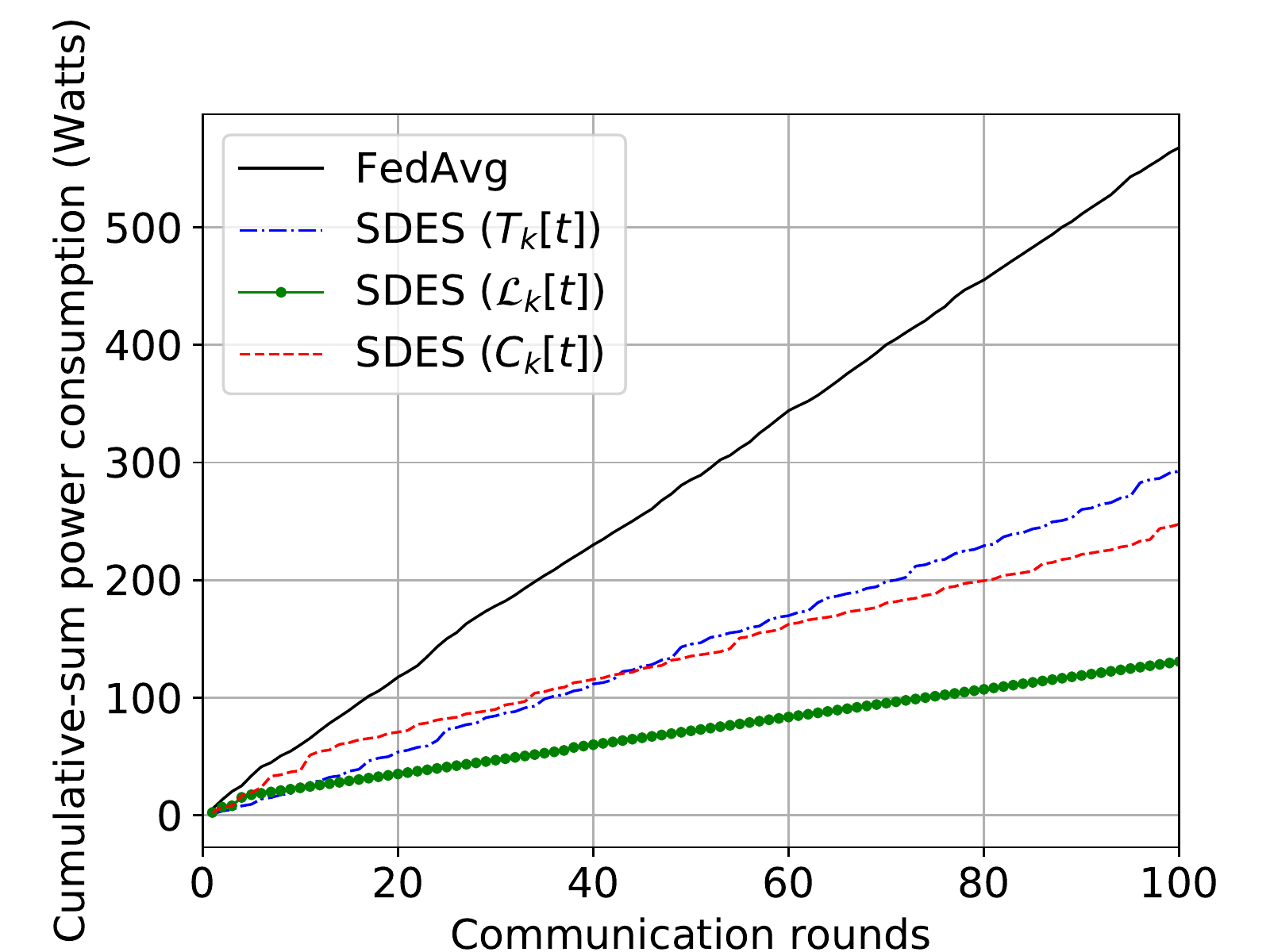}
	\end{minipage}}%
   \subfloat[Energy consumption ($W$=$K$)]{
		\begin{minipage}[t]{0.5\linewidth}
			\centering
			\includegraphics[width=3 in]{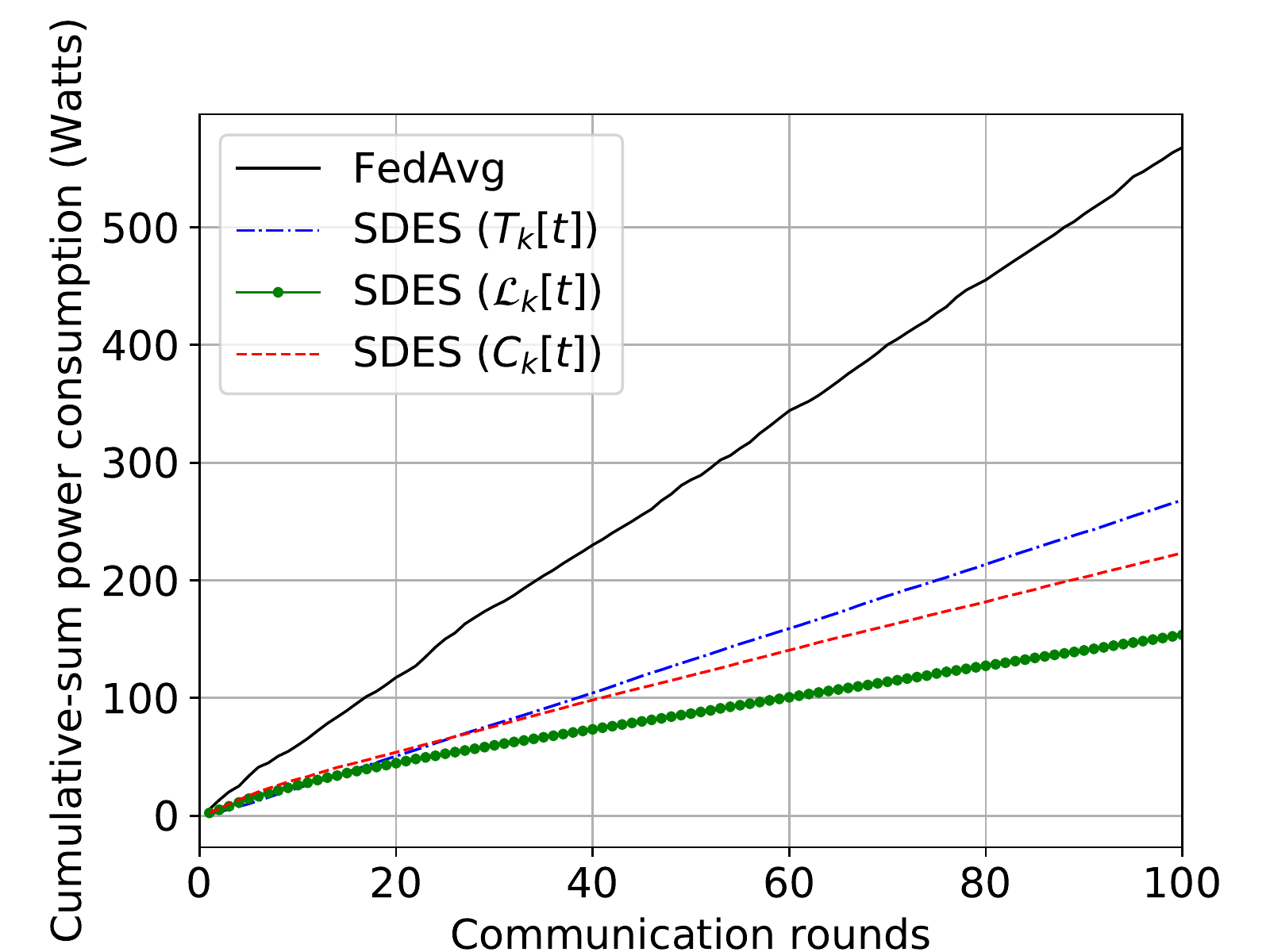}
	\end{minipage}}
	\\
	\subfloat[Model convergence]{
		\begin{minipage}[t]{0.5\linewidth}
			\centering
			\includegraphics[width=3 in]{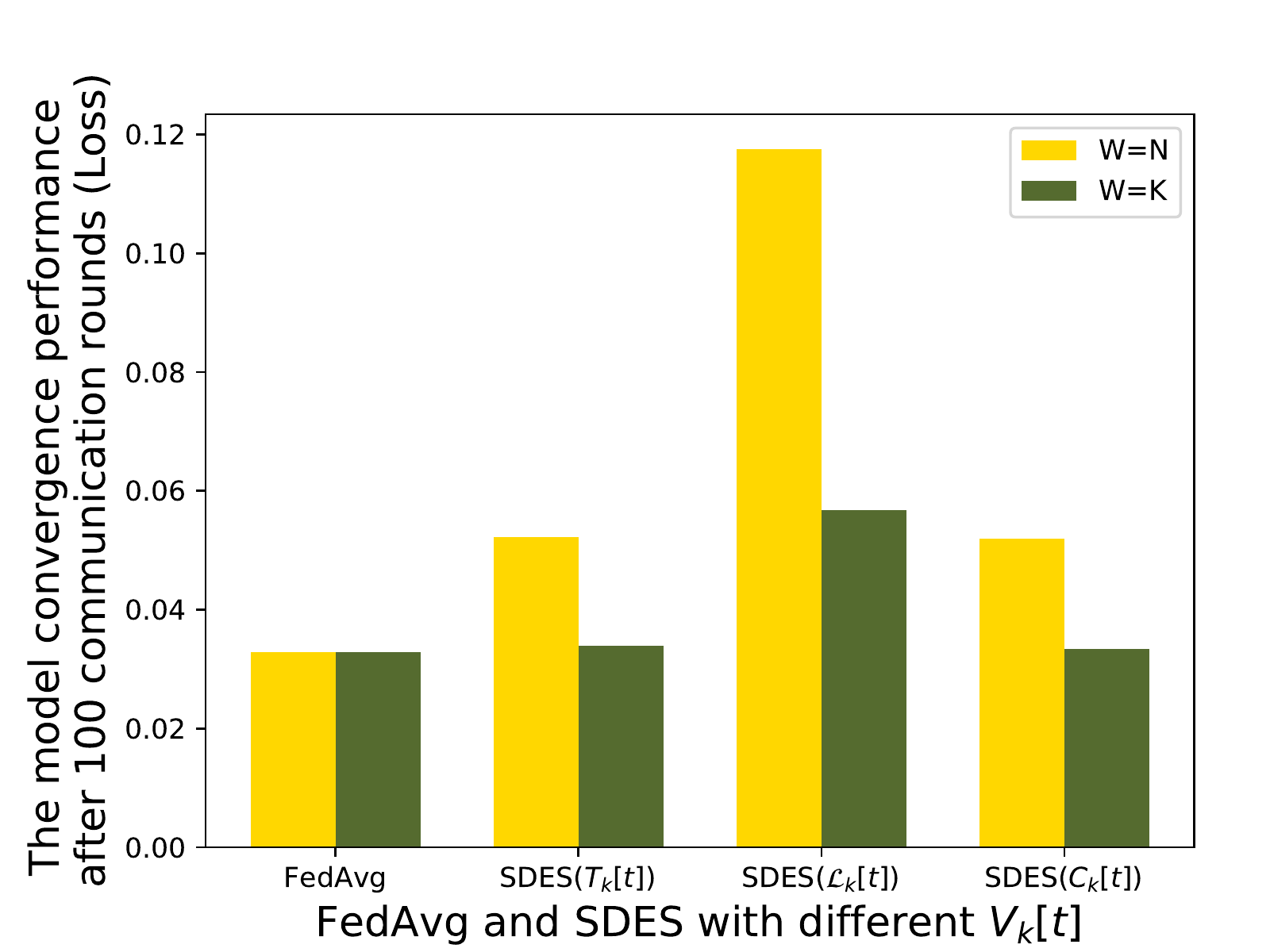}
   \end{minipage}}%
   \subfloat[Total energy consumption]{
		\begin{minipage}[t]{0.5\linewidth}
			\centering
			\includegraphics[width=3 in]{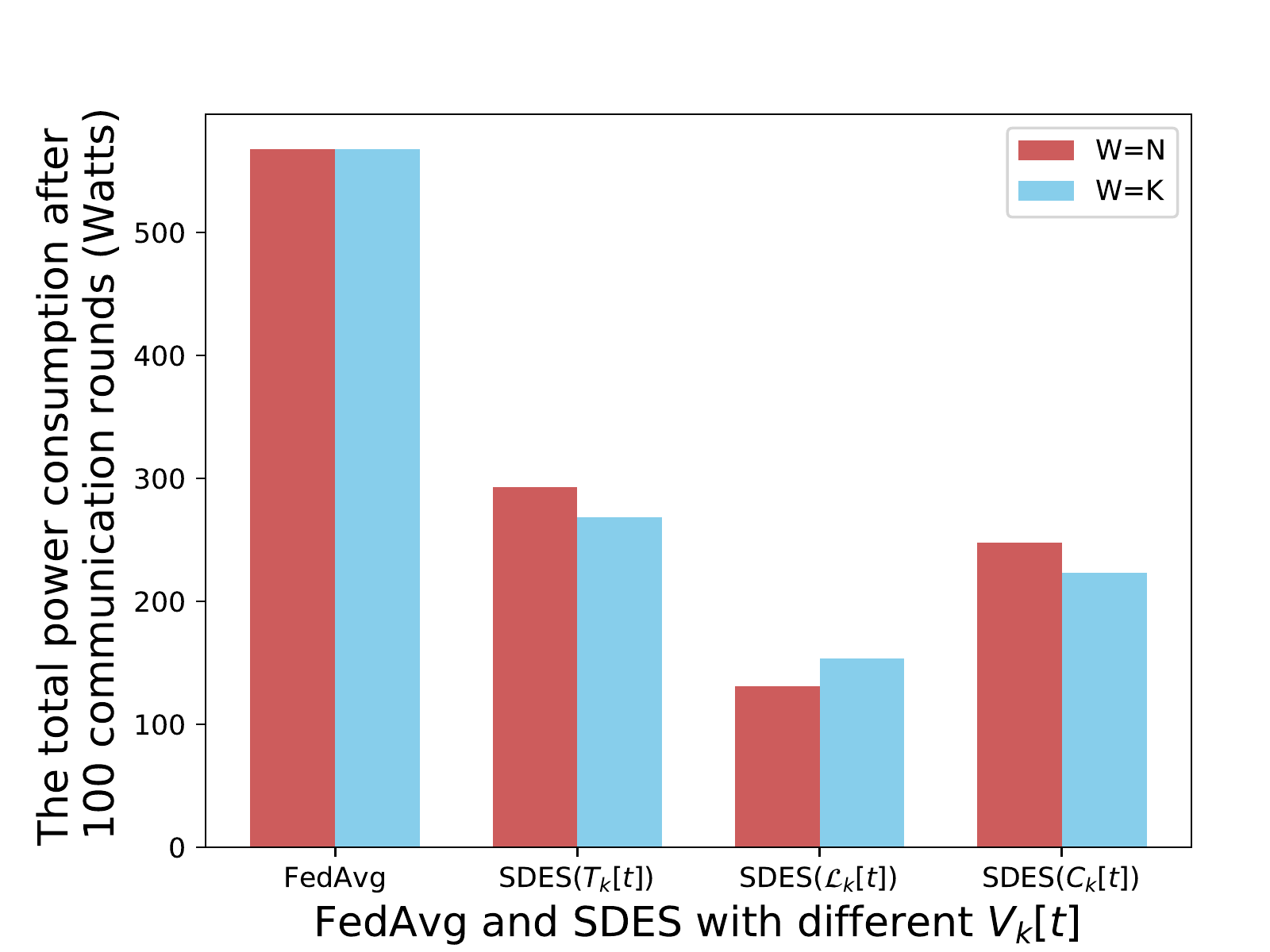}
	\end{minipage}}
	\caption{ SDES of $V_k[t] \in \{T_k[t], \mathcal{L}_k[t], C_k[t]\}$ ($\beta = 0.7, \ \zeta = 5$)}
	\label{4way-eta0.5}
\end{figure}

\begin{figure}[t]
	\subfloat[Instantaneous energy consumption]{
		\begin{minipage}[t]{0.5\linewidth}
			\centering
			\includegraphics[width=3 in]{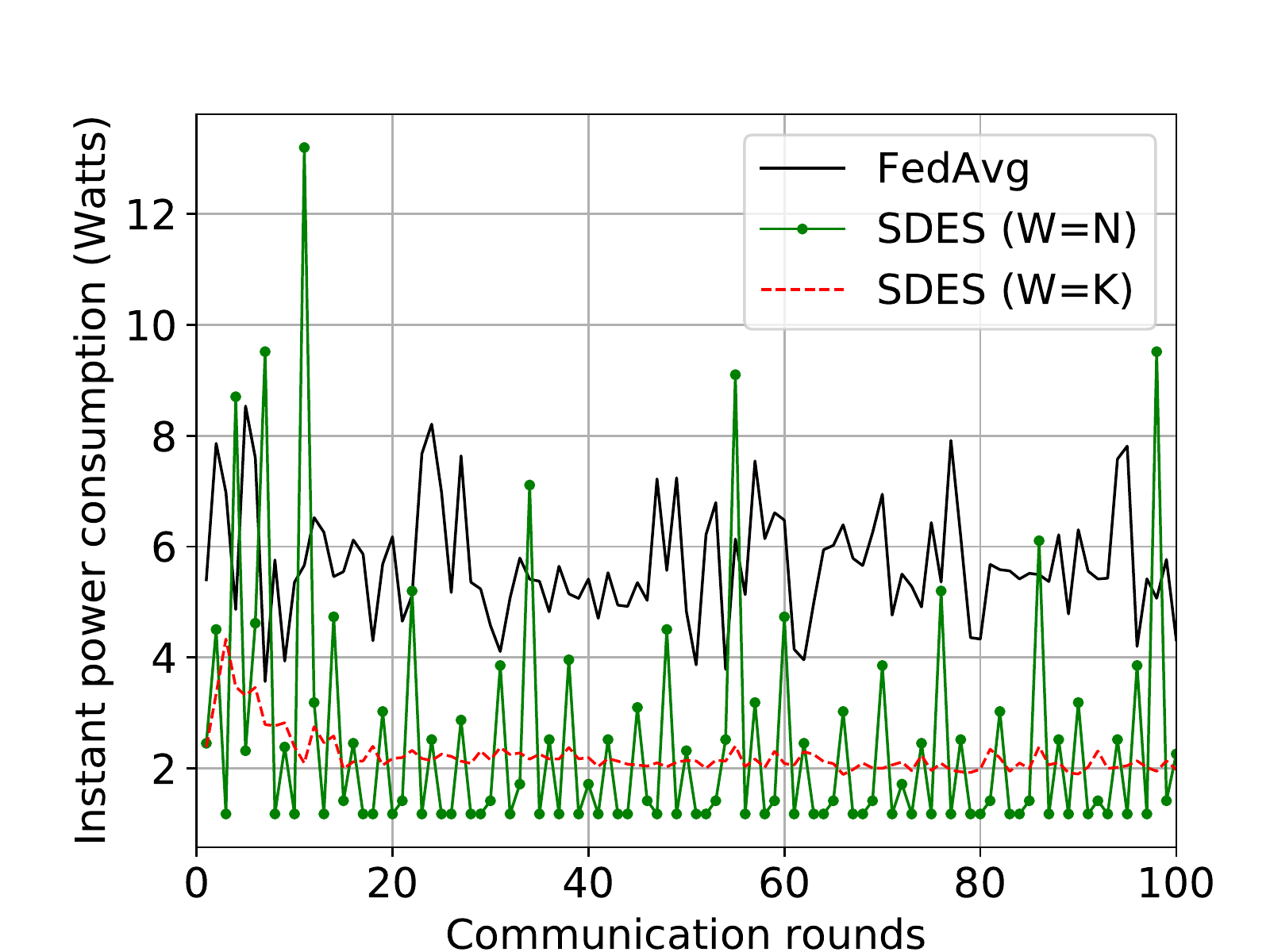}
	\end{minipage}} %
	\subfloat[Optimization function value \eqref{optim1}]{
		\begin{minipage}[t]{0.5\linewidth}
			\centering
			\includegraphics[width=3 in]{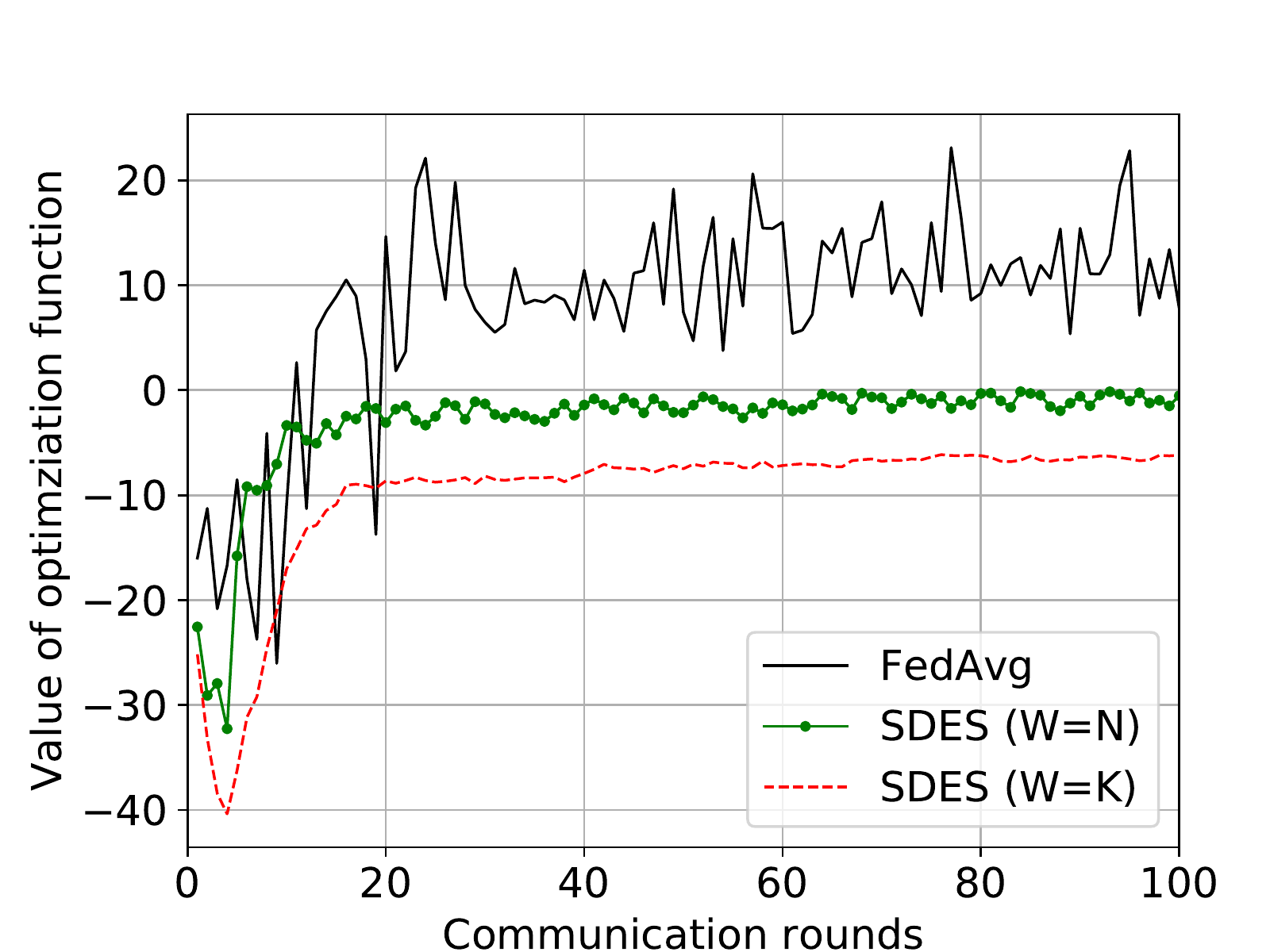}
	\end{minipage}}
	\caption{ SDES ($W \in \{K,N\}$) ($\beta = 0.7, \ \zeta = 5,\ V_k[t]=C_k[t]$)}
	\label{combined-0.5}
\end{figure} 

Fig.~\ref{4way-eta0.5} shows the performance gain of the proposed $C_k[t]$ measure in \eqref{Combined_Cov}, compared with staleness $T_k[t]$ and training loss $\mathcal{L}_k[t]$. One can see that 
in Fig.~\ref{4way-eta0.5}(a) and (b), SDES with $C_k[t]$ achieves good convergences similar to the optimal solution (i.e., FedAvg) which only considers the model convergence. FedAvg may often train the models of UEs with bad channel condition or with large dataset size, making it suffering from huge energy expense. Both of the cases converge fast at the beginning and also has good performance towards the end. 
In Fig. \ref{4way-eta0.5}(c) and (d), we compare the cumulative energy consumption among three measures. SDES with $\mathcal{L}_k[t]$ has the lowest cumulative energy consumption but with poor convergence performance, as shown in Fig. \ref{4way-eta0.5}(a). 
Moreover, SDES ($W$=$K$) and ($W$=$N$) with $C_k[t]$ have the first and second best performance in energy saving, as they consider both parameters of $\mathcal{L}_k[t]$ and $T_k[t]$.
Fig. \ref{4way-eta0.5}(e) and (f) shows that SDES achieves the best performance in energy conservation, and SDES with $C_k[t]$ and $W=K$ have the similar convergence performance as FedAvg in a more intuitive way.

Fig.~\ref{combined-0.5} further analyses the instant performance of SDES ($W$=$K$) and SDES ($W$=$N$) in terms of energy saving, where $C_k[t]$ in CR function is applied. One can see that both cases have good performance in energy saving compared with FedAvg. 
Moreover, one sees that the performances of SDES ($W$=$K$) with respect to model convergence in Fig. \ref{4way-eta0.5} and energy saving in Fig. \ref{combined-0.5}(a) are better than those of SDES ($W$=$N$). This is because high computational resource is required in case of ($W$=$K$), as explained in Section III.D.

The proposed SDES can be extended to more general cases, where the UEs are mobile with time-varying channels or several APs are deployed in FL. In the former case, the energy consumption of UEs is constantly changing. In the latter case, UEs send the trained models to the appropriate APs considering the channel condition, and APs then centralize all the data into one AP for the global model training. Compared with the benchmark solution of FedAvg, SDES bears acceptable computational complexity in the real-time application. Moreover, the choice of weight factor should be careful, since bad choice may lead to unacceptable model convergence performance.

\section{Conclusion}
In this paper, we have proposed a novel energy-efficient scheduling policy, i.e., SDES for federated learning in bandwidth-limited systems with energy-limited UEs. We have utilized the CR function for model convergence and introduced the SDES algorithm, which can reduce the computational complexity with parallel computing architecture. Simulation shows that our proposed SDES performs well in model convergence, and it can save energy consumed by UEs significantly compared with the benchmark solution in bandwidth-limited networks.
In the future, we will focus on the energy efficiency in the more practical federated learning cases in wireless communication, where the dataset contains complicated real information and the UE size is extended to thousands scale.

\bibliographystyle{IEEEtran}
\bibliography{references}

\begin{thebibliography}{10}
\providecommand{\url}[1]{#1}
\csname url@samestyle\endcsname
\providecommand{\newblock}{\relax}
\providecommand{\bibinfo}[2]{#2}
\providecommand{\BIBentrySTDinterwordspacing}{\spaceskip=0pt\relax}
\providecommand{\BIBentryALTinterwordstretchfactor}{4}
\providecommand{\BIBentryALTinterwordspacing}{\spaceskip=\fontdimen2\font plus
\BIBentryALTinterwordstretchfactor\fontdimen3\font minus
  \fontdimen4\font\relax}
\providecommand{\BIBforeignlanguage}[2]{{%
\expandafter\ifx\csname l@#1\endcsname\relax
\typeout{** WARNING: IEEEtran.bst: No hyphenation pattern has been}%
\typeout{** loaded for the language `#1'. Using the pattern for}%
\typeout{** the default language instead.}%
\else
\language=\csname l@#1\endcsname
\fi
#2}}
\providecommand{\BIBdecl}{\relax}
\BIBdecl

\bibitem{park2019wireless}
J.~Park, S.~Samarakoon, M.~Bennis, and M.~Debbah, ``Wireless network
  intelligence at the edge,'' \emph{Proceedings of the IEEE}, vol. 107, no.~11,
  pp. 2204--2239, 2019.

\bibitem{luo2020power}
Y.~Luo, J.~Yang, W.~Xu, K.~Wang, and M.~Di~Renzo, ``Power consumption
  optimization using gradient boosting aided deep q-network in c-rans,''
  \emph{IEEE Access}, vol.~8, pp. 46\,811--46\,823, 2020.

\bibitem{lu2018mimo}
C.~Lu, W.~Xu, H.~Shen, J.~Zhu, and K.~Wang, ``{MIMO} channel information
  feedback using deep recurrent network,'' \emph{IEEE Communications Letters},
  vol.~23, no.~1, pp. 188--191, 2018.

\bibitem{8845636}
C.~{Lu}, W.~{Xu}, S.~{Jin}, and K.~{Wang}, ``Bit-level optimized neural network
  for multi-antenna channel quantization,'' \emph{IEEE Wireless Communications
  Letters}, vol.~9, no.~1, pp. 87--90, 2020.

\bibitem{mcmahan2016communication}
H.~B. McMahan, E.~Moore, D.~Ramage, S.~Hampson \emph{et~al.},
  ``Communication-efficient learning of deep networks from decentralized
  data,'' \emph{arXiv preprint arXiv:1602.05629}, 2016.

\bibitem{zhao2018federated}
Y.~Zhao, M.~Li, L.~Lai, N.~Suda, D.~Civin, and V.~Chandra, ``Federated learning
  with non-iid data,'' \emph{arXiv preprint arXiv:1806.00582}, 2018.

\bibitem{FEDL}
C.~Dinh, N.~H. Tran, M.~N. Nguyen, C.~S. Hong, W.~Bao, A.~Zomaya, and
  V.~Gramoli, ``Federated learning over wireless networks: Convergence analysis
  and resource allocation,'' \emph{arXiv preprint arXiv:1910.13067}, 2019.

\bibitem{yang2019scheduling}
H.~H. Yang, Z.~Liu, T.~Q. Quek, and H.~V. Poor, ``Scheduling policies for
  federated learning in wireless networks,'' \emph{IEEE Transactions on
  Communications}, 2019.

\bibitem{zeng2019energy}
Q.~Zeng, Y.~Du, K.~K. Leung, and K.~Huang, ``Energy-efficient radio resource
  allocation for federated edge learning,'' \emph{arXiv preprint
  arXiv:1907.06040}, 2019.

\bibitem{AoU}
H.~H. Yang, A.~Arafa, T.~Q. Quek, and H.~V. Poor, ``Age-based scheduling policy
  for federated learning in mobile edge networks,'' \emph{arXiv preprint
  arXiv:1910.14648}, 2019.

\bibitem{chen2019deep}
J.~Chen and X.~Ran, ``Deep learning with edge computing: A review,''
  \emph{Proceedings of the IEEE}, vol. 107, no.~8, pp. 1655--1674, 2019.

\bibitem{storn1996usage}
R.~Storn, ``On the usage of differential evolution for function optimization,''
  \emph{Proceedings of North American Fuzzy Information Processing}, pp.
  519--523, 1996.

\bibitem{opara2019differential}
K.~R. Opara and J.~Arabas, ``Differential evolution: A survey of theoretical
  analyses,'' \emph{Swarm and evolutionary computation}, vol.~44, pp. 546--558,
  2019.

\end{thebibliography}
\end{document}